\documentclass{iau} 
\usepackage{graphicx}

\title[Morpho-Kinematics around Post-AGB Stars] 
{Morpho-Kinematics of the Circumstellar Environments around Post-AGB Stars}

\author[T.\ Ueta, H.\ Izumiura, I.\ Yamamura, \& M.\ Otsuka]   
{Toshiya Ueta$^1$,
Hideyuki Izumiura$^2$,
Issei Yamamura$^3$, \&
Masaaki Otsuka$^4$}

\affiliation{$^1$University of Denver, Denver, CO 80112, U.S.A.,
email: {\tt toshiya.ueta@du.edu} (TU) \\[\affilskip]
$^2$Okayama Branch Office, Subaru Telescope, 
NAOJ, Okayama, Japan,\\
$^3$Institute of Space and Aeronautical Science, 
JAXA, Kanagawa, Japan,\\
$^4$Okayama Observatory, Kyoto University, Okayama, Japan}

\pubyear{2018}
\volume{343}  
\jname{Why Galaxies Care About AGB Stars: A Continuing Challenge through Cosmic Time}
\editors{F.\ Kerschbaum, M.~A.~G.\ Groenewegen, \& H.\ Olofsson, eds.}
\begin{document}

\maketitle

\begin{abstract}
We observed two proto-planetary nebulae, HD\,56126 representing a source with an elliptical circumstellar shell and IRAS~16594$-$4656 representing a source with a bipolar circumstellar shell, with ALMA in the $^{12}$CO and $^{13}$CO J$=$3$-$2 lines and neighboring continuum to see how the morpho-kinematics of CO gas and dust emission properties in their circumstellar environments differ.

\end{abstract}

\firstsection 

\section{Dual Morphologies of Proto-Planetary Nebulae}

Proto-planetary nebulae (PPNe) are post-asymptotic giant branch (post-AGB) stars surrounded by a physically detached circumstellar envelope (CSE), which resulted from mass loss during the AGB phase up to about $10^{-4}$\,M$_{\odot}$\,yr$^{-1}$ (\cite{kwok1993,vanwinckel2003}).
PPNe are known to possess either the elliptical or bipolar morphology that appears to be developed by the beginning of the PPN phase, 
which is punctuated by the cessation of mass loss, supposedly because of the equatorially-enhanced mass loss during the final epochs of mass loss along the AGB (\cite{m02,u03}).

\section{ALMA Observations of Proto-Planetary Nebulae}

\begin{figure}[h]
\begin{center}
 \includegraphics[height=2.2in]{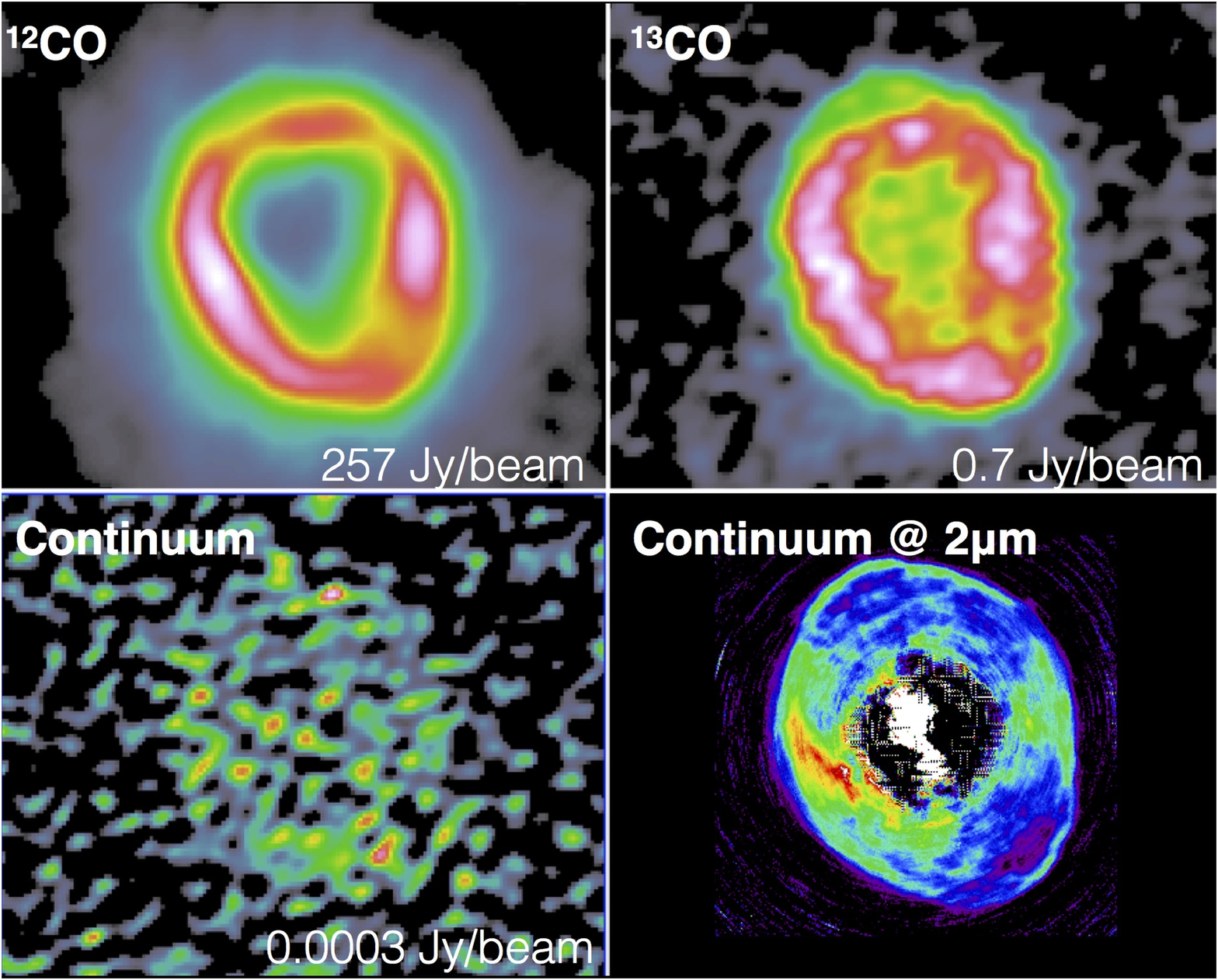} 
 \includegraphics[height=2.2in]{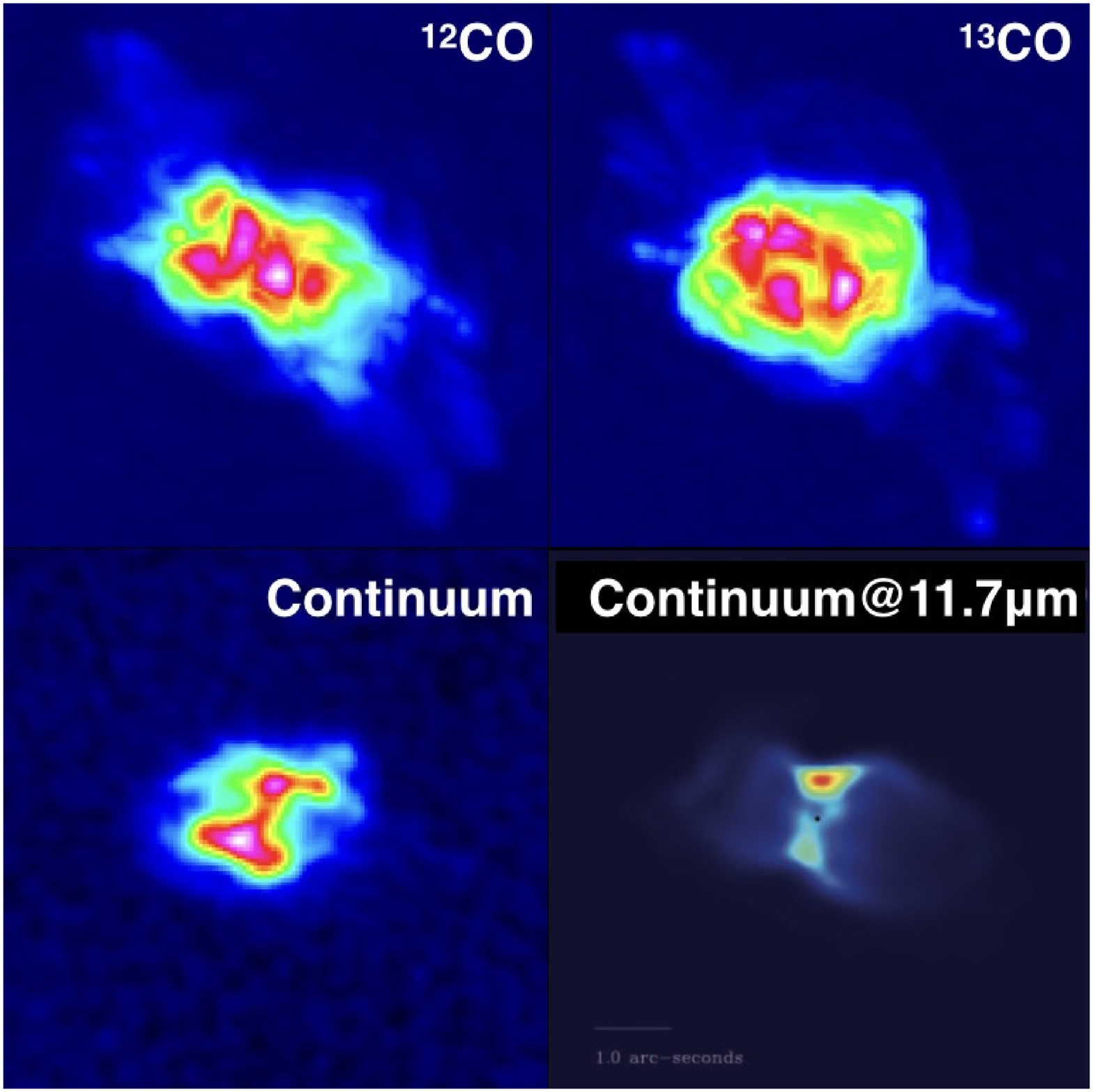} 
 \caption{%
 ALMA integrated maps in the $^{12}$CO, $^{13}$CO, and continuum, with a comparison continuum map in the IR for
 HD\,56126 (left panels; the shell diameter is $\sim$4$^{\prime\prime}$; the comparison 2\,$\mu$m map by Ueta (2015)) and 
 IRAS\,16594$-$4656 (right panels; the FoV is roughly $8^{\prime\prime}\times8^{\prime\prime}$).}
   \label{fig1}
\end{center}
\end{figure}

We observed two PPNe,
HD\,56126 and IRAS\,16594$-$4656, each of which represents one of these two morphological archetypes, elliptical and bipolar, respectively,
with the ALMA 12-m array in Band 7 during Cycle 3.
These PPNe were observed 
in the $^{12}$CO and $^{13}$CO J=3-2 lines
at the effective velocity resolution of 26 and 885\,m\,s$^{-1}$,
respectively, in 3840 channels.
We also observed these sources in the neighboring continuum bands
centered at 333.0 and 343.3\,GHz at the velocity resolution of 27.2\,km\,s$^{-1}$
in 128 channels.
The spatial resolution in terms of the beam size was 
$0.4^{\prime\prime} \times 0.3^{\prime\prime}$ and 
$0.5^{\prime\prime} \times 0.4^{\prime\prime}$,
respectively.
As for data reduction, we adopted the pipeline calibration performed 
with the Common Astronomy Software Application (CASA; ver.\ 4.5.3)
as delivered and made use of the newer CASA (ver 5.1.1-rel5) to 
use the new {\sc tclean} function with the 3-$\sigma$ threshold masking
to clean each velocity channel map during the final image reconstruction.
The use of the new {\sc tclean} function was necessary to clean 
algorithmically 800 and 3000 channels at which spatially-variable extended CO emission was detected
from HD\,56126 and IRAS\,16594$-$4656, respectively.
Fig.\,\ref{fig1} shows the integrated maps of HD\,56126 (left panel) and IRAS\,16594$-$4656 (right panel)
in the $^{12}$CO, $^{13}$CO, neighboring radio continuum, and continuum in the IR for comparison (from top-left to bottom-right), respectively.

\section{Morpho-Kinematics of Proto-Planetary Nebulae}

Thermal dust emission at 330--340\,GHz is almost non-existent for HD\,56126 but shows 
some concentration at the bipolar waist (i.e., dust torus) seen edge-on which resembles what was detected
 at 11.7\,$\mu$m \cite{v06} for IRAS\,16594$-$4656, 
 corroborating the optically-thin and thick nature of each of these dust shells, respectively.

HD\,56126 is essentially a hollow shell expanding at $\sim$8\,km\,s$^{-1}$ with a slight velocity enhancement
along the pole/long axis (at $\sim$12\,km\,s$^{-1}$).
There are two velocity components at the pole at $\sim$8 and $\sim12$\,km\,s$^{-1}$, indicating 
the presence of a separate, expanding cavity at the tips of the pole.
IRAS\,16584$-$4656 exhibits much more complex structures consisting of the central main bipolar structure
surrounding the dust torus and at least four pairs of highly elongated but much fainter lobes 
at $\sim$25\,km\,s$^{-1}$
(three in the plane of the sky and one almost perpendicular).
The morpho-kinematics is clearly distinct in these PPNe 
and is being analyzed to constrain the driving mechanism(s).

\firstsection 

\acknowledgment
This paper makes use of the following ALMA data: ADS/JAO.ALMA\#2015.1.00441.S. 
ALMA is a partnership of ESO (representing its member states), NSF (USA) and NINS (Japan), together with NRC (Canada), MOST and ASIAA (Taiwan), and KASI (Republic of Korea), in cooperation with the Republic of Chile. The Joint ALMA Observatory is operated by ESO, AUI/NRAO and NAOJ.
The National Radio Astronomy Observatory is a facility of the National Science Foundation operated under cooperative agreement by Associated Universities, Inc.
TU acknowledges technical assistance from the NA ARC at the NAASC.
Dr.\ Daniel Tafoya at the EA ARC is especially thanked for his assistance in data reduction.
In addition, TU appreciates fruitful discussions on the interpretation of the data with Dr.\ Griet Van de Steene at the Royal Observatory of Belgium.

\firstsection 

\end{document}